\documentclass{aa} 
\usepackage{times,amssymb,amsmath} 
\usepackage{epsfig}

\title{Detection of Lyman-$\alpha$ emission from a DLA
galaxy: Possible implications for a luminosity-metallicity relation
at $z$ = 2--3
\thanks{Based on observations collected at the European Southern
Observatory, Paranal, Chile (ESO Programme 66.A-0386(A)}
}

\author{P. M\o ller \inst{1}
        \and J.P.U. Fynbo \inst{2,3}
         \and S.M. Fall \inst{4}
        }

\institute{
           European Southern Observatory,  
           Karl-Schwarzschild-Stra\ss e 2,
	   D-85748, Garching by M\"unchen, Germany
           \and
	   Department of Physics and Astronomy, University of {\AA}rhus,
	   Ny Munkegade, 8000 {\AA}rhus C, Denmark
	   \and
	   Astronomical Observatory, University of Copenhagen, 
	   Juliane Maries Vej 30, 2100 Copenhagen {\O}
           \and
           Space Telescope Science Institute, 3700 San Martin Drive,
           Baltimore, MD 21218
           }
\offprints{}
\mail{pmoller@eso.org}

\date{Received  / Accepted }

\begin{document}
\titlerunning{Ly$\alpha$ emission from the $z=2.0395$ DLA towards
PKS0458-02}

\abstract{
In an ongoing programme to identify a sample of high $z$ DLA galaxies we
have found the long sought for case of a Ly$\alpha$ emitter seen in the
centre of a broad DLA trough. This is the predicted ``textbook case''
of an intervening DLA galaxy if DLA galaxies are small, but would not
be expected if intervening high redshift DLA galaxies have large
gaseous disks. The Ly$\alpha$ flux is
$\rm 5.4^{+2}_{-0.8}\times 10^{-17} ergs~ s^{-1} cm^{-2}$
similar to what has been found
in previously known high $z$ DLA galaxies. The impact parameter
is found to be $0\farcs3\pm 0\farcs3$. This is smaller than what
was found in previous cases but still consistent with random sight-lines
through absorbers with mean impact parameter $\approx 1$\arcsec .
Of the 24 DLAs targeted in the NICMOS imaging survey five have now
been identified as Ly$\alpha$ emitters.
The DLA galaxies with detected Ly$\alpha$ emission tend to have
higher interstellar metallicities than those with undetected
Ly$\alpha$ emission. This is plausibly explained as a consequence 
of a positive correlation between the Ly$\alpha$ line luminosities 
of the galaxies and their metallicities, although the present 
sample is too small for a definitive conclusion. The available
observations of high-redshift DLA galaxies are also consistent 
with a negative correlation between Ly$\alpha$ equivalent widths and 
metallicities, as seen in nearby star-forming galaxies and
usually attributed to the preferential absorption of Ly$\alpha$ 
photons by dust grains.
\keywords{galaxies: formation - galaxies: high-redshift - quasars:
absorption lines - quasars: individual (PKS 0458-02)
}
}

\maketitle

\section{Introduction}
Shortly after publication of the first sample of Damped Ly$\alpha$
Absorbers (DLAs, Wolfe et al. 1986) it was suggested that Ly$\alpha$
emission from the absorbing galaxies should be detectable as a central
spike in the trough of the damped Ly$\alpha$ line (Foltz et al. 1986).
Two conflicting views on the nature of high redshift DLA galaxies
were in disagreement on this prediction. Wolfe (1986) and Smith et al.
(1986) both
suggested that the 5 times higher cross-section of high-redshift DLAs 
relative to local spirals indicated that DLAs were fully formed 
disks with radii $\sim\sqrt{5}$ times larger than locally. Tyson (1988),
on the other hand, argued that DLAs could be gas-rich dwarf galaxies.
Two observing strategies were adopted reflecting the two views.
Under the assumption that the DLA is small its impact parameter
relative to the QSO shall also be
small, and therefore a long-slit centred on the QSO should have a high
probability of also covering the 
DLA (e.g. Hunstead et al. 1990). If on the other 
hand DLAs are large disks they may have large impact parameters relative
to the QSOs and narrow band imaging would be a better approach
(Smith et al. 1989; Wolfe et al. 1992).

At redshifts $\leq 1$ photometric redshifts are now known for 11 DLA
galaxies (Chen \& Lanzetta 2003) but evolutionary models predict that
DLA galaxies at higher redshifts are of a different nature
(Lanfranchi \& Friaca 2003).

In this paper we present the detection of Ly$\alpha$ emission from the 
z=2.0395 DLA towards PKS0458-02 at $z=2.286$.
This detection was obtained
in the course of a spectroscopic investigation of three candidate DLA
galaxy counterparts reported by Warren et al. (2001). The Ly$\alpha$
emission from the DLA towards PKS0458-02 does not correspond to any
of the candidates at projected distances 0\farcs86-4\farcs17,
on the contrary it is
found almost exactly in the centre of the DLA absorption line, thereby
presenting itself nicely as a ``textbook example'' of Ly$\alpha$
emission from a high-redshift DLA.

\section{Observations and data reduction}
Long-slit spectroscopy of PKS0458-02 was obtained during two dark
nights at the ESO Very Large Telescope in October 2000. We used FORS1
with the G600B grism and a 1\farcs31 wide slit providing a spectral
resolution of about 7\AA \ FWHM for objects wider
than the slit. During read-out we binned the data 2 by 2 providing
final pixels of size 0\farcs4 by 2.22\AA . The 
conditions during observations were photometric and the seeing 
ranged from 0\farcs6 to 0\farcs9 resulting in a resolution of
4-5\AA \ for point sources.  A total integration time of 
11400 seconds was split as follows: 
Three 2000 second and one 1400 second exposures at a slit position angle
of +55\fdg0 (East of North); two 2000 second exposures at a slit
position angle of +28\fdg8. The two position angles were chosen to
be lined up with three candidate galaxies N-6-1D, N-6-4C, and N-6-5C
(Warren et al. 2001). The individual spectra were bias subtracted
following standard techniques, but after extensive testing we found
that the spectroscopic daytime flats were of inferior quality and
that better flat fielding was obtained with a combination of a U band
imaging flat and a 1D ``along slit efficiency curve'', both obtained
during twilight. Acquisition images were taken in R-Bessel and we
found R(Bess) = $18.43\pm0.05$ for PKS0458-02.

\section{Results}

We did not detect emission lines in any of the three targeted candidates
(at impact parameters 0\farcs86, 2\farcs98, and 4\farcs17) but at both
PAs we clearly detect an emission line in the centre of the DLA
absorption line, and at close to zero impact parameter (Fig.~1). Since
this object does not appear in any candidate list
we shall in what follows name it DLAg0458-02.

\begin{figure}
\vskip -0.5cm
\begin{center}
\epsfig{file=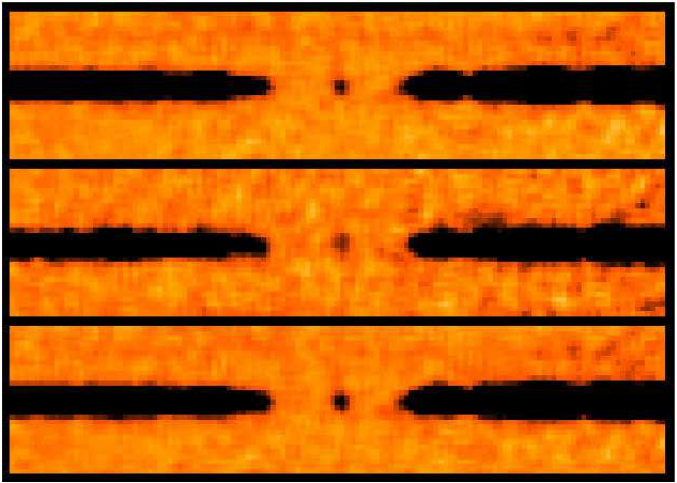, width=6.4cm, angle=270}
\caption{2D spectra of the DLA at $z=2.0395$. The top spectrum is
obtained with the slit at PA=55\fdg0 (E of N), the middle
spectrum at PA=28\fdg8, the bottom spectrum is the weighted
sum of the two. In all three blue is to the left red to the right,
upwards is roughly NE. The emission peak is clearly seen in the
centre of the broad absorption feature.}
\end{center}
\vskip -0.2cm
\end{figure}

\begin{figure}
\begin{center}
\epsfig{file=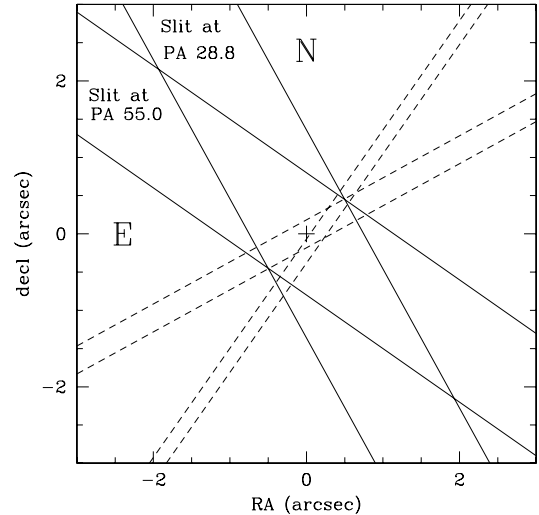, width=7cm}
\vskip -0.3cm
\caption{Layout of the slits (full drawn lines) and determined
1$\sigma$ ranges of the position of the DLA galaxy DLAg0458-02
(dashes). The position of PKS0458-02 is marked by the ``+''.}
\end{center}
\end{figure}

At PA=55\fdg0 and PA=28\fdg8 we find impact parameters of
$b_{55}$=0\farcs13$\pm$0\farcs09 and 
$b_{28.8}$=0\farcs00$\pm$0\farcs16 respectively. In Fig.~2
we show the layout of the slit positions (solid lines) and the
1$\sigma$ ranges of the position of DLAg0458-02 inferred from the
two independent detections. In principle two different slit PAs is
enough to triangulate for the exact position of the DLA galaxy, but
unfortunately the two PAs are rather close to each other so there is
still uncertainty about the exact impact parameter.

\begin{figure}
\vskip -0.5cm
\begin{center}
\epsfig{file=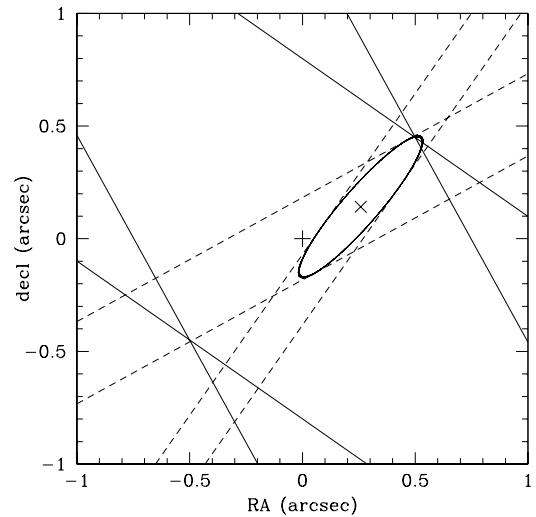, width=7cm}
\vskip -0.3cm
\caption{Inner 2\arcsec\ by 2\arcsec\ of the field around PKS0458-02.
The ellipse marks the combined 1$\sigma$ contour of the position
of DLAg0458-02. The most likely position is marked by an ``$\times$'' at
PA=300$^o$, $b_{\rm DLA}$=0\farcs30.}
\end{center}
\vskip -0.5cm
\end{figure}

In Fig.~3 we show an enlargement of the inner 2\arcsec\ by 2\arcsec\
where we
have calculated the combined 1$\sigma$ contour. The best fit position
of DLAg0458-02 is at PA=300$^o$, $b_{\rm DLA}$=0\farcs3, but the entire
PA range from 175$^o$ to 315$^o$ is allowed to within 1$\sigma$. The
full 1$\sigma$ range on $b_{\rm DLA}$ is from 0\farcs0 to 0\farcs6.
There is a hard upper limit of $b_{\rm DLA} \leq 0\farcs8$ as a larger
impact parameter would cause the object to fall outside the slit. We
conclude that $b_{\rm DLA}=0\farcs3\pm 0\farcs3$.
An impact parameter of $b=0\farcs3$ is too small for a detection in
the NICMOS survey of Warren et al. (2001).

\begin{figure}
\vskip -0.3cm
\begin{center}
\epsfig{file=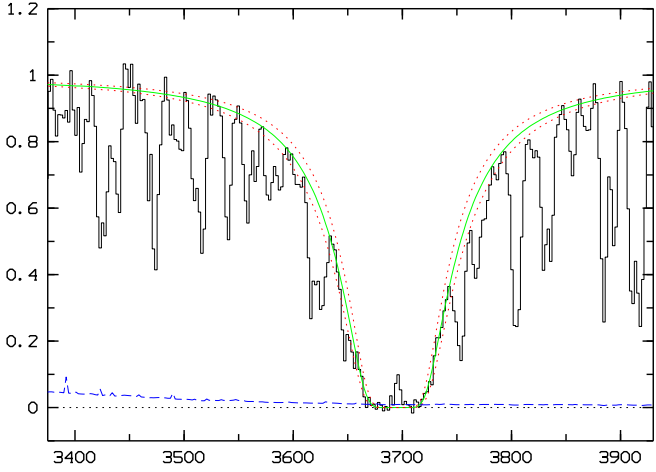, width=11cm, angle=0}
\vskip -9cm
\caption{FORS1/G600B spectrum of the $z=2.0395$ Damped Ly$\alpha$ line.
Calculated Ly$\alpha$ lines with columns of
      $6\times 10^{21} {\rm cm}^{-2}$  (full green line),
5 and $7\times 10^{21} {\rm cm}^{-2}$ (dotted red lines) are also
shown. Ly$\alpha$ emission from the DLA galaxy is seen in the centre
of the absorption line. The 1$\sigma$ noise spectrum (per 5\AA\
resolution element) is shown as a dashed line (blue). The
emission line is detected at 9.3$\sigma$.}
\end{center}
\vskip -0.1cm
\end{figure}

In Fig.~4 we show the 1D extraction of the spectrum (optimally weighted
sum of both PAs) and the Ly$\alpha$ emission line is clearly seen in
the centre of the Damped absorption line. We determine the centroid of
the emission line to be at 3696.86\AA \ corresponding to a redshift of
2.0410 which is 148~km~s$^{-1}$ higher than that of the DLA. However,
this is only correct if one assumes $b_{\rm DLA}=0$. For our best fit
value of $b_{\rm DLA}=0\farcs3$ we must correct the redshift for the
offset of the object inside the slit and we then obtain a redshift of
2.0396 which is identical to that of the DLA. We shall adopt this latter
redshift as the best fit value. The emission line is unresolved at our
resolution and we find an upper limit of FWHM $\leq 5$\AA \
corresponding to an upper limit of 400~km~s$^{-1}$ FWHM.

As seen in Fig.~4 the absorption profile is well fitted by a damped
Ly$\alpha$ line with an HI column density of
$6 \pm 1 \times 10^{21} {\rm cm}^{-2}$. This is slightly larger
than, but consistent with, the values found by
Wolfe et al. 1993 ($5\times 10^{21} {\rm cm}^{-2}$) and
Pettini et al. 1994 ($4.5\pm 1 \times 10^{21} {\rm cm}^{-2}$ ).

The flux of the Ly$\alpha$ emission line, as well as other relevant
measurements, are summarised in Table 1. The listed $+1\sigma$ on the
Ly$\alpha$ flux is mostly due to the uncertainty of the correction
for slit-losses (which are linked to the error on $b_{\rm DLA}$),
the listed $-1\sigma$ is mostly due to the uncertainty of the flux
calibration. The Ly$\alpha$
flux of DLAg0458-02 is within the range found for Ly$\alpha$
selected galaxies at the same redshift (Fynbo et al. 2002).
The continuum of the DLA galaxy is for the most part hidden under the
QSO spectrum but a small fraction of it could be visible in the
DLA trough. To place a limit on the equivalent width of the emission
line we proceeded as follows. On the best fitting Ly$\alpha$ absorption
line profile ($6 \times 10^{21} {\rm cm}^{-2}$) we determined the
interval in which the calculated absorption line was completely black
(i.e. residual flux below 1/5 of the rms in the data). This is the
case over an interval of 38\AA . The central
15\AA \ are taken up by the emission line, the rest we summed up to
see if a galaxy continuum could be detected. We did not detect any
continuum, and we find a 2$\sigma$ lower limit on the observed
equivalent width of 47\AA \ or a rest equivalent width of
$W_{\rm r}\geq 15.3$\AA \ ($2\sigma$ limit).

A long slit spectrum taken at PA=300$^o$ should be obtained in order to
verify the best fit impact parameter and PA found via triangulation.
This is needed to settle all open questions related to the exact
position of DLAg0458-02 inside the slit (redshift, velocity relative to
absorber, flux correction for slit loss).

\begin{table}
\caption{Data for PKS0458-02 and DLAg0458-02}
\begin{tabular}{lrl}
\hline
Quasar redshift                    & 2.286$^{[1]}$      & \\
Quasar R(Bessel)                   & $18.43\pm 0.05$      & \\
Absorption redshift                & 2.0395$^{[1]}$     & \\
N$_{\rm HI}$                  & $6\pm 1 \times 10^{21}$ & cm$^{-2}$ \\
Impact parameter, $b_{\rm DLA}$     & $0\farcs3\pm 0\farcs3$     & \\
Position angle           & $300^{+15}_{-125}$ & degrees (E of N) \\
Ly$\alpha$ emission line redshift  & 2.0396$^{[2]}$     & \\
$\Delta v$(Ly$\alpha$ - DLA)       & $10\pm150^{[2]}$   & km~s$^{-1}$ \\
Ly$\alpha$ FWHM                    & $\leq 400$         & km~s$^{-1}$ \\
Ly$\alpha$ flux&$5.4^{+2}_{-0.8}\times 10^{-17}$&$\rm ergs~ s^{-1} cm^{-2}$\\
Ly$\alpha$ EW$_{\rm rest}$         & $>15.3$ (2$\sigma$ limit) & \AA \\
\hline
\end{tabular}
$^{[1]}$ Pettini et al. 1994.
$^{[2]}$ After correction for inferred offset in slit.
\end{table}

\section{Discussion}
The Warren et al. (2001) NICMOS imaging survey targeted a sample of
24 DLAs and sub-DLAs ($N_{\rm HI} \la 2 \times 10^{20}$~cm$^{-2}$,
Dessauges-Zavadsky et al. 2003).
That sample was selected to cover a wide range in redshifts and in
HI column densities but absorber metallicities were not considered. At
the time of sample definition precise metallicities were known for
only very few DLAs and the sample therefore has no metallicity 
pre-selection biases. Of the 24 systems five have now been identified as
Ly$\alpha$ emitters. Three of the five have known metallicities while
for one additional system a lower limit to its
metallicity has been published. The metallicities and other relevant
information for the five identified systems is
summarised in Table 2. Metallicities are also known for
13 of the 19 systems for which no Ly$\alpha$ emission has yet been
reported (Kulkarni \& Fall 2002, Prochaska 2003, Dessauges-Zavadsky
et al. 2003).

\begin{table*}
\caption{Ly$\alpha$ emission
and metallicity data for five $z \geq 1.9$ DLA galaxies. $b_{\rm DLA}$
is the impact parameter, $log({\rm N})$ is the HI column density, and
$\Delta v$ is the difference between DLA redshift and QSO redshift
given in km~s$^{-1}$. The ``Type'' classifications in the last column
are defined as
``$z_a \approx z_e$'': $|\Delta v| \leq 3000$ km~s$^{-1}$; ~~
``sub-DLA'': $log({\rm N}) \leq 20.3$; ~~
``DLA'': Classic intervening non-sub DLAs. }
\begin{tabular}{cccclccllccc}
\hline
ID &\scriptsize NICMOS &$z_{\rm DLA}$ &$b_{\rm DLA}$ &emission
   &$log({\rm N})$ &[M/H] &M &[M/H] &$z_{\rm QSO}$ &$\Delta v$  &Type\\
   &\scriptsize ID     &              & \arcsec      &refs.
   &cm$^{-2}$      &      &  &refs. &              &km~s$^{-1}$ & \\
\hline
DLAg0528--25 & \scriptsize N-7-1C  & 2.8110 &1.14(2) & 1,3,5,10,11 &
 21.35 & -0.75,-0.76  & Si,Zn & 14,15 &2.797 &-1100 &$z_a \approx z_e$\\
sDLAg2233+13 & \scriptsize N-16-1D & 3.1493 &2.51(2) & 2,4,10,11,12&
 20.00 & $\geq -1.04$ & Si    & 16 &3.298 &10500 & sub-DLA\\
DLAg0151+04 & -       & 1.9342 &0.93    & 6,7,8,9     &
 20.36 &              &       &    &1.922 &-1200 &$z_a \approx z_e$\\
\hline
DLAg2206--19 & \scriptsize N-14-1C & 1.9205 &0.99(2) & 10,11       &
 20.65 & -0.39,-0.42  & Zn,Si & 15,17 &2.559 &58600 & DLA\\
DLAg0458--02 & -       & 2.0396 &0.3(3)  & 13          &
 21.78 & -1.17,-1.19  & Zn,Zn & 15,17 &2.286 &23300 & DLA\\
\hline
\end{tabular}
\\
{\footnotesize
(1) M{\o}ller \& Warren 1993;
(2) Steidel et al. 1995;
(3) Warren \& M{\o}ller 1996;
(4) Djorgovski et al. 1996;
(5) M{\o}ller \& Warren 1998;
(6) M{\o}ller et al. 1998;
(7) Fynbo et al. 1999;
(8) M{\o}ller 1999;
(9) Fynbo et al. 2000;
(10) Warren et al. 2001;
(11) M{\o}ller et al. 2002;
(12) Christensen et al. 2004;
(13) This paper;
(14) Lu et al. 1996;
(15) Kulkarni \& Fall 2002;
(16) Lu et al. 1998;
(17) Prochaska et al. 2003.
}
\end{table*}

\subsection{A possible luminosity-metallicity relation for DLA
galaxies}
Fig.~5a is a histogram of the interstellar metallicities of the 17
DLA systems in the Warren et al. (2001) NICMOS sample for which the
metallicity is known. The subsample of four objects with detected
Ly$\alpha$ emission is indicated by solid blue bars, while the 
subsample of 13 systems without detected Ly$\alpha$ emission is
indicated by red hatched bars. Evidently, there is a tendency for the
detected objects to have higher metallicities than the undetected
objects, although the sample is too small to draw a definitive
conclusion from this comparison. Four of the objects in the Warren
et al. sample might not be regarded as ``bona fide'' DLAs in the sense
that they are close to the QSO ($\Delta \la 3000$~km~s$^{-1}$) or that
they have relatively low HI column densities
($N_{\rm HI} \la 2 \times 10^{20}$~cm$^{-2}$).
Fig. 5b shows the metallicity histogram with these objects excluded.
Again, the two objects with detected Ly$\alpha$ emission have higher
metallicities, on average, than the 11 objects without detected
Ly$\alpha$ emission.

The trend displayed in Figs 5a and 5b is consistent with 
a positive correlation between Ly$\alpha$ line luminosity
and metallicity. This in turn is consistent with a correlation
between star formation rate, for which Ly$\alpha$ is an
approximate indicator, and metallicity. For nearby galaxies
($z = 0$), there are well known correlations between star
formation rate and mass and between metallicity and mass, in
the sense that high-mass galaxies tend to have both higher
star formation rates and metallicities than low-mass galaxies,
albeit with a great deal of scatter in samples that include
a wide or full mix of morphological types
(Tamura et al. 2001, Prada \& Burkert 2002).
If the same kinds of correlations also hold at high redshift, they
could explain the observed tendency for DLA galaxies with detectable
Ly$\alpha$ emission to have higher metallicities than those
without detectable Ly$\alpha$ emission. Conversely, the 
available observations of DLA galaxies could be interpreted 
as (weak) evidence for luminosity-metallicity and mass-metallicity
relations at $z \sim 3$. 

   The observed trend might, at first sight, appear to 
contradict the idea that Ly$\alpha$ emission is suppressed
in galaxies with high dust content and hence high metallicity.
Ly$\alpha$ photons, which scatter resonantly off H atoms,
have a much longer path and a much higher probability of
absorption on dust grains than continuum photons, an effect
that reduces the equivalent width $W_{\alpha}$ of the 
Ly$\alpha$ emission line. For nearby star-forming galaxies, 
the observed Ly$\alpha$ equivalent widths show the expected
anti-correlation with metallicity, which is usually
interpreted as a consequence of absorption by dust (see 
Fig. 8 of Charlot \& Fall 1993). For DLA galaxies at high 
redshifts, $W_{\alpha}$ has been measured in three cases and 
a lower limit placed on it in another case. We have checked 
that these equivalent widths and the corresponding 
metallicities are consistent with the $W_{\alpha}$-$Z$ 
relation for $z = 0$ galaxies and hence with the idea that the 
Ly$\alpha$ emission is suppressed by dust.

   How can we reconcile these two results, one regarding a
positive $L_{\alpha}$-$Z$ correlation, the other a negative
$W_{\alpha}$-$Z$ correlation? First, we note that the samples
on which these hints are based are still very small, and 
it is possible they will disappear with more observations.
Second, we note that there is no logical contradiction 
between the two correlations. They can both exist together,
as they are observed to do in low-redshift galaxies. The
reason this is possible is that dust affects $W_{\alpha}$ 
by factors of a few, while $L_{\alpha}$ spans several orders
of magnitude, from low-mass to high-mass galaxies. Thus, a
positive $L_{\alpha}$-$Z$ correlation can overwhelm a negative 
$W_{\alpha}$-$Z$ correlation. More observations are needed
to confirm whether similar relations also hold for high-
redshift DLA galaxies. For now, we simply note that the 
available but meager high-$z$ data are consistent with 
both low-$z$ relations.

\begin{figure}
\vskip -0.7cm
\begin{center}
\epsfig{file=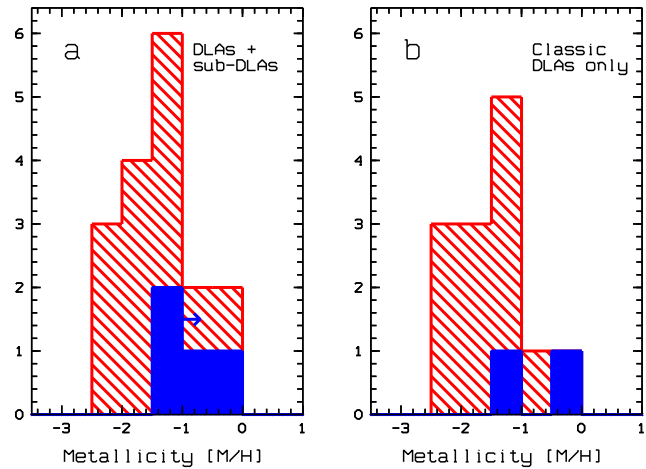, width=11.5cm}
\vskip -9.5cm
\caption{{\bf Left:} Histogram of metallicities of the 17 DLAs of the
NICMOS sample for which metallicities are known (hashed red). Of the 17
only 4 have been detected as Ly$\alpha$ emitters (shaded blue). Note
that for one of them the metallicity is a lower limit ($\geq -1.04$,
marked by the blue arrow) and may shift towards the right when a final
metallicity is determined. {\bf Right:} The same as histogram
(a) but
sub-DLAs and $z_{\rm abs} \approx z_{\rm em}$ DLAs have been excluded
leaving only 13 objects.}
\end{center}
\vskip -0.5cm
\end{figure}

\subsection{Sizes and impact parameters of DLAs}
The DLA absorber in front of PKS0458-02 was previously (Wolfe et al.
1985) detected as a 21 cm absorber. In particular, because PKS0458-02
has radio structure on a wide range of scales, Briggs et al. (1989)
were
able to use high-resolution radio interferometry to probe several
different paths through the absorbing medium. Using this technique
they concluded that the absorber is a disk-like structure that extends
across at least 2\arcsec . How does this earlier result fit with our
detection of an impact parameter of only $0\farcs3\pm 0\farcs3$?

A measured impact parameter is in each case random and
could be anywhere in the range $[0; R]$ where $R$ is the radius of the
absorber. M{\o}ller \& Warren (1998) found that in the mean the
measured impact parameter is $0.55R$, so finding a very small $b_{\rm
DLA}$ for DLAg0458-02 is therefore not in contradiction to the
result by Briggs et~al. Four optical (Ly$\alpha$) impact parameters
have been reported previously: 0\farcs93 (Fynbo et~al. 1999);
0\farcs99, 1\farcs14, and 2\farcs51 (M{\o}ller et~al. 2002). The
median of all five is 0\farcs99
corresponding to a disk diameter of 3\farcs6, fully
consistent with the Briggs et~al. result.

\subsection{Concluding remarks}
We shall not make any strong statement based on the above discussion
but simply point out that the data presented in Fig.~5 are certainly
consistent with, and are even weekly supporting,
the conjecture that a luminosity-metallicity relation is already in
place at redshifts 2-3. In further support of a
luminosity-metallicity relation we note that Lyman Break galaxies
(LBGs) have metallicities as high as or higher than the most metal rich
DLAs at similar redshifts (Pettini et al., 2000, 2001) but based on
cross-section selection arguments Fynbo et al. (1999) found that
typical DLAs have much lower luminosities than LBGs.

Detection of star formation induced Ly$\alpha$ emission from an
additional 5-10 DLAs would settle the questions raised above and should
therefore be a high priority observational goal. In order to
disentangle the dust-attenuation effect from the luminosity-metallicity
relation one would need a measure for the galaxy luminosity which is
not influenced by dust. This could be obtained via broad band imaging
of DLA galaxies with sufficiently large impact parameters, or via
detection of H$\alpha$/H$\beta$ emission lines.

If it is confirmed that a luminosity-metallicity relation
for DLA galaxies is indeed present then this might reflect an
underlying mass-metallicity relation.
We have already previously shown that a large fraction of DLA galaxies
are too small and too faint to be detected under the glare of the
quasar point spread function (Fynbo et al. 1999), but if
a mass-metallicity relation is indeed present already at high
redshifts then we can add to the previous statement that those
DLA galaxies most likely to be undetectable are those with the
lowest metallicities, a prediction which could greatly improve the
efficiency of follow-up observing campaigns. One further
observational prediction would
also follow immediately from this. Because the less massive DLA systems
have smaller radii we predict a correlation between impact parameter
and DLA metallicity.

\begin{acknowledgements}
We are grateful to S. J. Warren for comments on earlier versions of
this manuscript and to C. Ledoux for many helpful discussions
concerning metallicities of DLAs.
\end{acknowledgements}

\end{document}